\documentclass[aps,prab,groupedaddress,secnumarabic,amssymb,superscriptaddress]{revtex4}
\usepackage{amsmath,amssymb,amsfonts}
\usepackage[colorlinks]{hyperref}
\usepackage{graphicx}
\graphicspath{{./}}
\newcommand{\nn}{\nonumber}

\newcommand{\gev}{{\rm GeV}}

\newcommand{\mmmrad}{{\rm mm\cdot mrad}}

\newcommand{\temllmode}{${\rm TEM}_{11}$-mode\ }

\newcommand{\w}{\omega}

\newcommand{\re}{{\rm Re}\ }

\begin{document}

\title{THz deflector using dielectric-lined waveguide
for ultra-short bunch measurement
in GeV scale}%

\author{Shingo Mori}%
\email{smori@post.kek.jp}
\affiliation{KEK Accelerator department, Tsukuba, Ibaraki 305-0801, Japan}
\author{Mitsuhiro Yoshida}%
\email{mitsuhiro.yoshida@kek.jp}
\affiliation{KEK Accelerator department, Tsukuba, Ibaraki 305-0801, Japan}
\date{\today}%
\begin{abstract}
We propose an RF deflector in the THz regime
to measure the bunch length of
the ultrashort electron beam in GeV scale
by using the dielectric-lined circular waveguide (DLW) structure.
We show the design of the deflector and the possible resolution
in the attosecond scale with a reasonable input pulse energy of THz.
We investigate the short-range wakefield effect in the DLW to the time resolution using the analytical model based on the eigenmode calculation
and show the scaling law in terms of the
beam size, the bunch length.
We found the ideal resolution of the deflector can
reach order ${\cal O}(100)~[{\rm as}]$ with pulse energy of several ${\rm mJ}$ with a negligible wakefield effect.
Example calculations are given for a structure with
the vacuum hole radius of $0.5,\ 0.4,\ 0.3~[{\rm mm}]$,
the dielectric constant of $3.75$, the operating frequency of $0.2,\ 0.4,\ 0.6~[{\rm THz}]$, respectively.
\end{abstract}
\maketitle
\tableofcontents

\section{Introduction}

The longitudinal bunch-length measurement of the ultrashort electron beam is important
to optimize the performance of many electron-beam-based facilities such as free-electron lasers (FEL~\cite{emma2010first,ishikawa2012compact,kang2017hard}),
ultrafast electron diffraction (UED~\cite{chergui2009electron,sciaini2011femtosecond}), laser-driven plasma wakefield accelerator~\cite{esarey2009physics}, beam-driven plasma wakefield acceletor~\cite{litos2014high}, and the dielectric laser accelerator~\cite{england2014dielectric}.
In the FEL facilities, the RF deflector with a frequency of $6\sim11[{\rm GHz}]$
is widely used to measure the time information of the short electron bunch of $100~[{\rm fs}]$,
while the performance improvement requires a shorter bunch length less than $10~[{\rm fs}]$.
In such facilities, the required longitudinal bunch length is smaller than the resolution of the conventional RF deflectors.
Then, it is important to investigate the measurement method of
the longitudinal information of the ultrashort electron bunch.

The resolution of the RF deflector is linearly proportional to the deflecting voltage and
the operating frequency.
Due to the limitation of the input power and length of the deflecting tube, the higher operating frequency of the defector is required.
Among the available RF deflectors, ones with a frequency of $11~{\rm GHz}$  (X-band) have the best resolution, but in the microwave community, the higher frequency than X-band is not practically used due to the lack of power sources.

In both the accelerator and laser community, the development of the high energy and high peak-power THz source is investigated.
The laser-driven THz source was studied in the ref~\cite{ahr2017narrowband},
where narrowband THz radiation demonstrated by using periodically poled lithium niobate
(PPLN) crystals and driver pulses from a high-energy Ti:sapphire laser.
pulses from a high-energy Ti:sapphire laser
On the other hand, the electron-beam driven THz generation is also demonstrated in the ref~\cite{o2016observation},
where the circularly polarized THz is generated as a wakefield by
the intense relativistic electron beam decelerated by the field inside the dielectric-lined circular waveguide (DLW).
Due to the nature of the dielectrics, the DLW structure can support the high gradient of more than $1~[{\rm GV/m}]$.

The applications of the THz pulses to the accelerator components are investigated, e.g.
the THz-driven linear electron acceletor~\cite{wong2013compact,nanni2015terahertz,zhang2018segmented},
the electron gun~\cite{huang2016terahertz,fallahi2016short},
the undulator~\cite{curry2018meter}, the bunch compressor~\cite{wong2013compact,kealhofer2016all,zhao2018few,ehberger2019terahertz}, the streaking~\cite{kealhofer2016all,li2019terahertz,fabianska2014split,zhao2018terahertz,zhao2019terahertz,PhysRevAccelBeams.22.012803}.

We will investigate the possibility of THz deflector as a traveling wave tube of DLW supporting the linearly polarized THz multicycle pulse generated by the laser-driven THz source.
The DLW has the configuration of a metallic tube with an inner radius $b$, partially filled with dielectrics
with the relative dielectric constant $\epsilon_r$, which has a hole of radius $a$ as a beam hole.
This structure works similarly as an RF waveguide.
The linearly polarized THz and ultrashort electron bunch are injected into the DLW, where
the electron bunch passes the center of the beam axis.
The THz field propagates along the DLW as a \temllmode which has the deflecting electromagnetic field and
no longitudinal electric field on the beam axis.
If the phase velocity of the \temllmode of the DLW is $v_p=c$, the electron bunch stays in the same THz phase.

The THz deflector works the same as a usual RF deflector, which transforms the longitudinal information to the transverse
information by the deflecting kick to the bunch.
For the maximal resolution, the center of the electron bunch should set the phase of zero THz amplitude so that
the head and the tail of the bunch are kicked in the opposite direction and the center of the bunch keep on the axis without deflection.
Then, using the screen monitor set in the downstream from the deflector, we can read off the longitudinal bunch length
by measuring the transverse size of the deflected bunch.
To resolve the longitudinal bunch length, the kicked electrons should have a larger offset than the transverse bunch size without deflecting voltage.
As mentioned in the latter section, the resolution of the longitudinal length is determined by
the fraction of the transverse momentum given by the kick in the momentum of the bunch.

The THz deflector differs from the RF deflector in that
the dimension of the structure is determined by the THz wavelength, millimeter-scale, or less,
which is much smaller than the typical dimension of the RF deflector, several centimeters.
The smaller wavelength has an advantage in high energy density and high gradient.
However, the DLW with a small radius of the beam hole has large wakefield $\propto a^{-2}$
for a longitudinal wake and $\propto a^{-3}$ for a transverse wake, which
causes the beam instability and the energy spread.
In the THz deflector, the transverse offset of the bunch due to the alignment error is the source of
the transverse wake.
We will discuss the wakefield effect on the resolution of the longitudinal bunch length and the tolerance of
the alignment error.

In Sec.~\ref{sec:method},
we will describe the interaction between the injected THz
and the bunch in the DLW.
In Sec.~\ref{sec:wake},
we will introduce the analytical calculation of the
short-range wakefield for the ultrashort beam inside the DLW
with finite beam size and bunch length.
Sec.~\ref{sec:result} shows the designs and properties of the THz
deflector and the frequency error propagated from the fabrication error.
Here, we also show the wakefield contribution for the ultrashort
bunch in the DLW and fraction of the wakefield contribution compared
to the THz contribution and the scaling law of the
wakefield effect in terms of the beam size and the bunch length.
In Sec.~\ref{sec:conclusion}, we summarize the result.

\section{The contribution from the deflecting mode}\label{sec:method}

In this section, we introduce the momentum kick from the deflecting
mode to the bunch and the time resolution when the wakefield effect is
negligible.

The explicit formulae of the eigenmode are described in App.~\ref{sec:eigen}.
The deflecting force applied from the deflecting mode with
phase velocity $\beta_p$
to the particle with charge $q$
and velocity $\beta_b$ is given as follows:
\begin{eqnarray}
    F_r/q &=& \re \left( E_r^n - \beta_b c \mu_0 H_\phi^n \right),\nn\\
    &=&  - \gamma_p E_0
    R_n(\lambda_1,1-\beta_p\beta_b,(\beta_b-\beta_p)r_\eta)
    \cos(n\phi)\sin \psi(z,t),\\
    &\sim&-\frac{1}{2}\left((1-\beta_p \beta_b) + (\beta_b-\beta_p)r_\eta \right)
    \gamma_p E_0 \cos(n\phi)\sin \psi(z,t)
\end{eqnarray}
where we define $\gamma_p=1/\sqrt{1-\beta_p^2}$.
The radial dependence of the field is included in the dimensionless
function $R_n$ defined in Eq.~\eqref{eq:rn}.
$\lambda_1=\alpha_1 r$ is the dimensionless
radial coorinate.
In the third line, we approximate the field near the beam axis $\lambda_1\ll 1$.
$E_0$ is the maximum amplitude of the $E_z$ component.
$r_\eta=Z_0/\eta$ is the ratio of the
impedance in the free space $Z_0$ and
one for the propagating mode $\eta$ defined in Eq.~\eqref{eq:eta}.
Considering the synchronizing particle satisfying $\beta_b=\beta_p$,
we define the deflecting gradient $E_\perp$ as the amplitude of the deflecting field on axis,
$F_r/q\leq E_0/(2\gamma_p)\equiv E_\perp$.
For the synchronizing particle,
the contribution from the
axial component of the magnetic field proportional to $r_\eta$
vanishes,
which means the TEM mode can be seen as a pure TM mode for the particle.

Then the transverse momentum change $c\Delta p_\perp^{\rm RF}$ for the dipole mode
during the DLW of length $L_d$
can be calculated as follows:
\begin{eqnarray}
  c\Delta p_{\perp}^{\rm RF} &=& \int_0^{L_d}dz F_r,\\
  &\simeq&
  q_2 E_\perp^{\rm RF} L_d \gamma_p^2
  \left[
(1-\beta_p \beta_b) + (\beta_b-\beta_p)r_\eta
  \right]
  \left(k_b \zeta - \frac{1}{2}\delta k L_d\right),\label{eq:cprf}
\end{eqnarray}
where we define the wavenumber in free space $k_0=\omega/c$.
The last factor stems from the integration of the $(-\sin\psi(z,t))$ along
the longitudinal direction, where we use $\psi(z,t)=\omega t - k_z z$
and $z(t)=\beta_b c t + \zeta$ and the detail of calculation is given in Eq.~\eqref{eq:cptRFB2}.
We define the mismatch between the phase velocity $\beta_p$ and the bunch velocity $\beta_b$ as
$\delta k=(\beta_b^{-1}-\beta_p^{-1})k_0$ and the deflecting gradient of the
THz as $E_\perp^{\rm RF}$.
For the main deflecting mode, namely ${\rm TEM}_{11}$-mode,
$r_\eta\sim -1$.
In the second equality, we assume $\lambda_1,\ k_b \zeta,\ \delta k L_d\ll 1$.
The first term denotes the ideal deflecting voltage that applies to the particles in the bunch,
which is linear in terms of the displacement from the bunch center.
The second term comes from the mismatch between the phase velocity and the bunch velocity.
The effect of the velocity mismatch accumulates as the DLW becomes longer.

As described in App.~\ref{sec:resolution}, the time resolution of the
deflector is determined by the comparison of the transverse kick by the deflector
and the transverse rms size of the slice of the bunch measured at the screen at a distance of $L$
from the deflector.
As shown in the Eq.~\eqref{eq:resolution}, the time resolution of THz deflector
is given as
\begin{eqnarray}
  \Delta\zeta &>&  \frac{pc}{e V_\perp k_b\sin(\Delta \psi_x)}
  \frac{1}{\gamma_p^2((1-\beta_p \beta_b) + (\beta_b - \beta_p)r_\eta)}
  \frac{\epsilon_x}{x_{\rm rms}(s_0)\sin(\Delta \psi)}
  ,\label{eq:zetamin}
\end{eqnarray}
where $V_\perp=E_\perp^{\rm RF} L_d$, $k_b=k_0/\beta_b$
and $\epsilon_x$ is the horizontal geometrical emittance.
The right-hand side of Eq.~\eqref{eq:zetamin} is the minimum resolution in the measurement of the bunch length when the wakefield effect is negligible.

\section{The wakefield contribution to the deflector}\label{sec:wake}

If the passage of the bunch has a nonzero offset from the center line of the DLW,
the particle in the bunch is decelerated by the deflecting mode
and deflected by the wakefield
generated by the bunch itself.
As the deflection from the wakefield causes the transverse momentum change of $\Delta p_\perp^{\rm w}$,
we need to take into account the contribution in Eq.~\eqref{eq:resolution-condition} as
\begin{eqnarray}
  \Delta p_\perp(\zeta)= \Delta p_\perp^{\rm RF}(\zeta) + \Delta p_{\perp}^{\rm w}(\zeta),
\end{eqnarray}
where $\Delta p_\perp^{\rm RF}$ is the momentum change by the deflecting mode.
The magnitude of the RF contribution and the amplitude of the electric field strength
$E_0^{\rm RF}$ are determined by the input power of the THz, while
the amplitude of the wakefield $E_0^{\rm w}$ contribution is determined by
the energy loss of the bunch by the deceleration.

Following the discussion of the energy balance in reference~\cite{tremaine1997electromagnetic},
we can obtain the amplitude of the wakefield generated by each particle in the bunch in the
self-consistent manner.
In the following analysis, the quantity with tilde denotes one divided by electric field strength $E_0$ as $E_{z1}^{(nm)} = E_0^{(nm)} \tilde{E}_{z1}^{(nm)}$ and $U^{(nm)}=(E_0^{(nm)})^2 \tilde{U}^{(nm)}$,
where $U^{(nm)}$ is the electromagnetic energy per unit length for the propagating mode with
$n$ zimuthal waves and $m$ radial waves.
In the following formulae, the variables with superscript $(nm)$ are mode-dependent and calculated by using the
frequency of $nm$-th mode.

Due to the fundamental theorem of the beam-loading, the particle with charge $q_1~[{\rm C}]$ running during infinitesimal time $\delta t=\delta z/v_b$ is decelerated by ${\rm TEM}_{nm}$-mode (or ${\rm TM_{0m}}$-mode for $n=0$) and loses energy $\delta z q_1 E_z^{(nm)}(r_1,\phi_1)/2$.
Then the particle excites the mode in the region of $(v_b-v_g^{(nm)})\delta t$
behind the particle
, which has energy
$\delta z (1-v_g^{(nm)}/v_b)U^{(nm)}$,
where $v_g^{(nm)}$ is the group velocity of the $(nm)$-th mode.
Considering the energy conservation among the particle and field with negligible power loss during the propagation, the electric field strength of the mode in the wakefield can be obtained as a function of the transverse coordinate of the drive particle as
\begin{eqnarray}
  E_0^{(nm)}(r_1,\phi_1) &=& \frac{q_1 \tilde{E}^{(nm)}_z(r_1,\phi_1)}{2(1-v_g^{(nm)}/v_b)\tilde{U}^{(nm)}},\\
  &=& \frac{q_1}{2(1-v_g^{(nm)}/v_b)\tilde{U}^{(nm)}} I_n(\alpha_1 r_1) \cos{(n\phi_1)},
\end{eqnarray}
where the dimension of $\tilde{U}^{(nm)}$ is $[{\rm C\cdot m \cdot V^{-1}}]$.
Using the explicit formulae shown in Eq.~\eqref{eq:uem}, we can calculate $\tilde{U}^{(nm)}$ as
\begin{eqnarray}
    \tilde{U}^{(nm)} = \frac{1}{2cZ_0}
    \left(\frac{1}{(\alpha_1^{(nm)})^2}\Lambda_{u1}^{(nm)} + \frac{1}{(k_2^{(nm)})^2}\Lambda_{u2}^{(nm)}\right),
\end{eqnarray}
where $\Lambda_{u1}^{(nm)}$ and
$\Lambda_{u2}^{(nm)}$ are the dimensionless integral
and requires to integrate numerically for each mode.
Then we can calculate the transverse force that the drive particle $q_1$ applies
to the trailing particle $q_2$ as follows:
\begin{eqnarray}
F_r^{(nm)}
&=&
q_2\gamma_p E_0^{(nm)}c_{n2}
R_n\left(
\alpha_1^{(nm)}r_2, 1- \beta_p \beta_b, (\beta_b - \beta_p)r_\eta
\right)(-\sin{(k_z^{(nm)} s)})
\\
&=&
\frac{q_1 q_2}{2(1-v_g^{(nm)}/v_b)\tilde{U}^{(nm)}}\frac{1}{\gamma_b}I_n(\alpha_1^{(nm)} r_1)
I_n'(\alpha_1^{(nm)} r_2)c_{n1}c_{n2}(-\sin{(k_z^{(nm)} s)}),
\end{eqnarray}
where we define the transverse coordinates of the particle $q_i$ as $\vec{r}_i=(r_i,\phi_i)$ and the temporal coordinates as $z_i(t) = \zeta_i + v_b c t$.
For simplicity we use $c_{ni}$ to abbrebiate $\cos{(n\phi_i)}\ (i=1,2)$.
$\zeta_i$ denotes the axial position measured from the bunch center, which is assumed unchanged during the bunch passing the DLW.
In the second equality, we assume the phase velocity of the wakefield satisfies
$\beta_p=\beta_b$.
The phase is the difference between the phase of the two particle
$\Delta \psi(s)=\psi_2(z_2,t_2)-\psi_1(z_1,t_1)=k_z^{(nm)}s$, where
$s=\zeta_1-\zeta_2$.

Considering the interaction between two electrons with charge $e$ inside the bunch,
the momentum change of the following electron caused by $nm$-th mode is
$c\Delta p_{\perp}^{(nm,{\rm 2e})}= L_d F_r^{(nm)},$
where we assume the distance between the two electrons $s$ is constant during the bunch
passing the DLW.

Considering the Gaussian bunch of cylindrical shape passing the transverse coordinate $(r_0,\phi_0)$ with the normalized charge distribution
\begin{eqnarray}
\rho_t(t,t_b)&=&\frac{1}{\sqrt{2\pi}\sigma_t}\exp{
\left(
-\frac{(t+t_b)^2}{2\sigma_t^2}
\right)
},\\
\rho_r(r,\phi)&=&\frac{1}{2\pi \sigma_r^2}\exp{
\left(
-\frac{1}{2\sigma_r^2}(r^2+r_0^2-2r_0 r \cos{(\phi - \phi_0)})
\right)
}
\end{eqnarray}
with the bunch legth $\sigma_t=\sigma_z/c$ for the time direction and the beam size $\sigma_r$ for
the transverse direction,
we can take into account the effect of the electron distribution
inside the bunch
parametrized by $\sigma_r$, $\sigma_t$, $r_0$.
Without loss of generality, we can choose $\phi_0=0$.
Then we can calculate
the momentum kick of the electron at $\zeta_2=c t_2$ from the leading electrons in the bunch, $\Delta p_\perp^{(nm,{\rm eG})}(\tau_2=\zeta_2/\sigma_z)$ as
\begin{eqnarray}
  c\Delta p_\perp^{(nm,{\rm eG})}(\tau_2) &=&
  N_e\int^{\infty}_{0}dt\rho_t(t+t_2,0)
  \int d\Pi_2 c\Delta p_\perp^{(nm,2e)}(r_1,\phi_1,r_2,\phi_2,ct)
  ,\\
  \int d\Pi_2 &=& \Pi_{i=1}^2\int^a_0 r_i dr_i \int^{2\pi}_0 d\phi_i \rho_r(r_i,\phi_i),
\end{eqnarray}
where $d\Pi_2$ denotes the measure of the integral in terms of the transverse coordinates $(r_i,\phi_i)$ with
the Gaussian weight $\rho_r(r_i,\phi_i)$ for $i=1,2$.

Therefore, we obtain
\begin{eqnarray}
  c\Delta p_\perp^{(nm,{\rm eG})}(\tau_2) &=&
  -
  \frac{eQL_d}{2(1-v_g^{(nm)}/v_b)\tilde{U}^{(nm)}}\frac{1}{\gamma_b}
  F_e\left(\tau_2,k_z^{(nm)} \sigma_z\right)G_n(\alpha_1^{(nm)}\sigma_r,r_0/\sigma_r).\label{eq:cpeg}
\end{eqnarray}
To factor out the effect of the
finite bunch length $\sigma_z=c\sigma_t$
and the beam size $\sigma_r$,
we define the dimensionless form-factors $F_e$
for the bunch length and $G_n$ for the beam size as follows:
\begin{eqnarray}
  F_e(\tau_2,k_z^{(nm)}\sigma_z) &=&
  \frac{1}{\sqrt{2\pi}}\int^\infty_{0}
  d\tau
  \exp\left(-\frac{(\tau_2+\tau)^2}{2}\right)
  \sin{(k_z^{(nm)}\sigma_z\tau)},\\
  G_n(\alpha_1^{(nm)}\sigma_r,r_0/\sigma_r)
  &=&\int^{a/\sigma_r}_{0} d\varrho_1 \int^{a/\sigma_r}_{0}d\varrho_2
  \varrho_1 \varrho_2
  e^{-\frac{1}{2}(\varrho_1^2+\varrho_2^2+2\varrho_0^2)}I_n(\varrho_1 \varrho_0)
  I_n(\varrho_2 \varrho_0)\nn\\
  &&I_n(\alpha_1^{(nm)} \sigma_r \varrho_1)
  I_n'(\alpha_1^{(nm)} \sigma_r \varrho_2),
\end{eqnarray}
where $\varrho_i=r_i/\sigma_r\ (i=0,1,2)$
and $D=\{(x,y)|x^2+y^2\leq a^2\}$, $x_0=r_0\cos\phi_0$, $y_0=r_0\sin\phi_0$.
Apparently, the radial form-factor $G_n$ takes into account the
loss of the particles which are outside the region of the vacuum hole.

The wakefield contribution is described by the sum of the all
eigenmodes excited by the bunch.
We define the net wakefield contribution to the bunch
as
\begin{eqnarray}
  c\Delta p_{\perp}^{\rm (eG)}(\tau) &=&
  \sum_{n=0}^N\sum_{m=1}^Mc\Delta p_{\perp}^{(nm,{\rm eG})}(\tau).\label{eq:cpteG-sum}
\end{eqnarray}
Here we introduce the UV cutoff from the bunch length namely $k_{0}^{(nm)}\sigma_z<1$,
as the higher-order mode with smaller wavelength than the bunch generated by one particle
in the bunch is canceled by ones generated by other particles and cannot accumulate as
the wakefield contribution from the net Gaussian bunch.
Then we define the maximum radial index $M$ as the largest number to satisfy
$k_{0}^{(nM)}\sigma_z<1$ for each $n$.
As for the azimuthal index $n$, the series expansion
provides the suppression factor for large $n$ as
$I_n(\alpha_1 r) I_n'(\alpha_1 r)   = {\cal O}((\alpha_1 r)^{2n-1})$ for
$n\geq1$, then we will consider $N=1$.

\section{Result}\label{sec:result}

\subsection{The designs of THz deflector and the time resoluiton}

Considering the charged particle with the
relativistic parameter $\gamma_b=1+E_e/(m_ec^2)=10^4$
and the relative permittivity of $\epsilon_r=3.75$,
the propagating constant $k_z$ relates to the wavelength in the free space $k_0$ as
$k_z=k_0/\sqrt{1-1/\gamma_p^2}$ with we assume $\gamma_p=\gamma_b$.
Then we obtain the dimension paramters $a$ and $r_{ab}=a/b$ as shown in Tab.~\ref{tab:dimension}
for the operating frequency of $f_0=200,\ 400,\ 600~[{\rm GHz}]$.

\begin{table}[htb]
  \begin{tabular}{cccccccc}\hline
    $\gamma_b$ & $\epsilon_r$ & $a~[{\rm mm}]$ & $r_{ab}$ & $f_0~[{\rm GHz}]$ & $Z_\perp[{\rm M\Omega/m}]$ & $v_g^{(11)}/c$ & $N_{\rm cycle}$\\ \hline
             &              & $0.5$          &  $0.7613$    & $200.2$ & $0.37$ & $0.565$ & $5.13$\\
    $10^4$   & $3.75$       & $0.4$          &  $0.85978$ & $400.04$ & $0.25$ & $0.702$ & $5.66$\\
             &              & $0.3$          &  $0.879845$& $600.025$ & $0.34$ & $0.74$ & $7.03$\\ \hline
  \end{tabular}\centering
  \caption{The dimension of the DLW which has \temllmode synchronizing
  with the particle of $\gamma_b=10^{4}$ with the hole radius of $a=0.5,\ 0.4,\ 0.3~[{\rm mm}]$ corresponding to
  the operating frequency of $f_0=200,\ 400,\ 600~[{\rm GHz}]$, respectively. The transverse shunt impedance per unit length $Z_\perp$ and the group velocity $v_g^{(11)}/c$ are calculated by the formulae shown in Eq.~\eqref{eq:vg} and Eq.~\eqref{eq:zsh}. $N_{\rm cycle}$ is the minimum number of THz cycle to deflect the particle along the DLW with length of $L_d=10~[{\rm mm}]$.}\label{tab:dimension}
\end{table}

Besides the dimension of the DLW, we show the result of the group velocity and
the transverse shunt impedance per unit length in Tab.~\ref{tab:dimension}.
The group velocity $v_g=P_z/U$
and the shunt impedance per unit length $Z_\perp=V_\perp^2/(P_z L_d)$
are obtained as
\begin{eqnarray}
  v_g/c &=& 2\left[\gamma_p^4\Lambda_{p1}
  +\left(\frac{1}{\sqrt{\beta_p^2\epsilon_r-1}}\right)^4\Lambda_{p2}\right]
  \left[\gamma_p^2\Lambda_{u1}^n
  +\left(\frac{1}{\sqrt{\beta_p^2\epsilon_r-1}}\right)^2\Lambda_{u2}^n\right]^{-1},\label{eq:vg}\\
  Z_\perp L_d &=& \frac{Z_0}{4}(k_0L_d)^2\frac{1}{\gamma_p^2\beta_p^2}
  \left[
  \gamma_p^4\Lambda_{p1}^n + \frac{1}{(\beta_p^2\epsilon_r- 1)^2}\Lambda_{p2}^n
  \right]
  .\label{eq:zsh}
\end{eqnarray}
The detail of our notation used in the above equations is described in App.~\ref{sec:eigen}.
Given the group velocity, we can calculate the number of cycle of the
THz without passed by the bunch with energy $\gamma_b=10^4$ as
$N_{\rm cycle} = f\left(\frac{L_d}{v_g}-\frac{L_d}{v_b}\right),$
where we use $L_d=10~[{\rm mm}]$.

Given the injected THz with energy of $10~[{\rm mJ}]$ and the pulse length of $10$ cycles,
the DLW designed in Tab.~\ref{tab:dimension} provides the diflecting gradient of
$E_\perp=86.1,\ 100,\ 143~[{\rm MV/m}]$ for $f_0=200,\ 400,\ 600~[{\rm GHz}]$, respectively.

The fluctuation of the frequency $\delta f/f_0$ can be expanded
in terms of the fluctuation of the radius of the vacuum hole $\delta a/a_0$, the radius of the metal boundary $\delta b/b_0$,
and the relative permittivity $\delta \epsilon_r/\epsilon_{r0}$ as below.
\begin{eqnarray}
  -\frac{\delta f}{f_0} = C_a^f \frac{\delta a}{a_0} + C_b^f \frac{\delta b}{b_0} + C_\epsilon^f \frac{\delta \epsilon_r}{\epsilon_{r0}},
\end{eqnarray}
where $c_a^f$, $c_b^f$ and $c_\epsilon^f$ are the coefficients obtained by numerical calculation.
Considering the typical error of the inner diameter and the
outer diameter of the
dielectric tube is $2\delta a=2\delta b = 0.01~[{\rm mm}]$,
we can estimate $(\delta f/f_0)\sim 2.4,\ 5.2,\ 8.0~[\%]$ for
$a=0.5,\ 0.4,\ 0.3~[{\rm mm}]$ corresponding to $f_0=200,\ 400,\ 600~[{\rm GHz}]$, respectively.

\begin{table}[htb]
  \begin{tabular}{cccccc}\hline
    $a~[{\rm mm}]$ & $f_0~[{\rm GHz}]$ & $C_a^f$ & $C_b^f$ & $C_\epsilon^f$\\\hline
    $0.5$ & $200$         & $1.00026$ & $3.2084$ & $0.450412$ \\
    $0.4$ & $400$         & $1.00041$ & $4.81607$ & $0.357122$ \\
    $0.3$ & $600$         & $1.00073$ & $5.43963$ & $0.334888$ \\\hline
  \end{tabular}\centering
  \caption{The error-propagation coefficients of the frequency of \temllmode in DLW
   in terms of the paramters $a$, $b$, and $\epsilon_r$ for the combinations
   for $a=0.5,\ 0.4,\ 0.3~[{\rm mm}]$ corresponding to $f_0=200,\ 400,\ 600~[{\rm GHz}]$, respectively,
   as shown in Tab.~\ref{tab:dimension}.}\label{tab:fluctuation}
\end{table}

As shown in Eq.~\eqref{eq:zetamin}, if the frequency tuning of
the injected THz
can match the phase velocity as $\beta_p=\beta_b$,
the time resolution becomes
\begin{eqnarray}
  \Delta\zeta &>&  \frac{cp}{e V_\perp k_b}
    \frac{\epsilon_x}{x_{\rm rms}(s_0)\sin(\Delta \psi_x)},
\end{eqnarray}
where $\epsilon_x$ denotes
the geometrical emittance
in the horizontal direction
and $x_{\rm rms}(s_0)$ is the beam size at the deflector $s=s_0$,
and $\Delta \psi_x$ denotes the phase advance between the deflector and the screen. $cp$ denotes the electron energy entering the THz deflector.
$V_\perp$ denotes the deflecting voltage and $k_b=k_0/\beta_b$, where $k_0$ is the wavenumber of the THz.

\begin{figure}[!htb]
  \centering
  \includegraphics*[height=5cm, bb= 0 0 367 233]{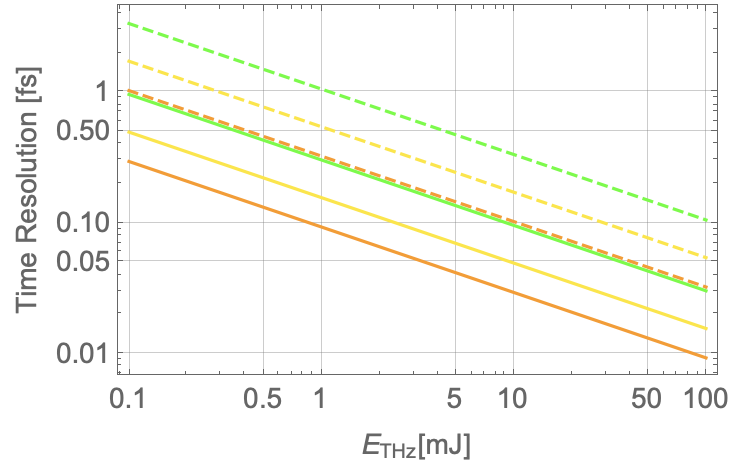}
  \caption{The time resolution of the THz deflector in terms of the pulse energy $E_{\rm THz}~[{\rm mJ}]=(N_{\rm cycle}V_\perp^2)/(f^{(11)} Z_\perp L_d)$ with
  the vacuum-hole size $a=0.3,\ 0.4,\ 0.5~[{\rm mm}]$ as shown in Tab.~\ref{tab:dimension} for the orange, yellow, and green lines,
  respectively.
  The solid and dashed lines denote the
  beam energy $cp=2.3,\ 8~[{\rm GeV}]$, respectively
  Here we assume $N_{\rm cycle}=10$, $L_d=10~[{\rm mm}]$, and
  the beam size $x_{\rm rms}(s_0)=a$.
  }
  \label{fig:resolution}
\end{figure}

Fig.~\ref{fig:resolution} shows the time resolution of the DLW
for the synchronizing particle $\beta_p=\beta_b$.
Assuming typical normalized emittance $\epsilon_x^n = 0.1~[\mmmrad]$ and $\Delta \psi_x = \pi/2$, and the beam energy of $cp=2.3,\ 8~[\gev]$,
we obtain the plot of
the time resolution as shown in Fig.~\ref{fig:resolution}
for the vacuum hole radius $a=0.3,\ 0.4,\ 0.5~[{\rm mm}]$.
Here we assume the rms beam size at the deflector is chosen as large
as the vacuum hole radius of the DLW, $x_{\rm rms}(s_0)=a$, to have larger
time resolution,
and the deflecting voltage is
calculated as $V_\perp^{\rm RF}=\sqrt{E_{\rm THz}f^{(11)}Z_\perp L_d /N_{\rm cycle}}$ assuming the length of the DLW is $L_d=10~[{\rm mm}]$ and
the number of injected THz cycle per pulse $N_{\rm cycle}=10$.
Then the DLW potentially has fine time resolution less than $0.1~[{\rm fs}]$
even the THz pulse energy of $E_{\rm pulse}=10~[{\rm mJ}]$.

\begin{figure}[!htb]
  \centering
  \includegraphics*[height=5cm, bb= 0 0 353 244]{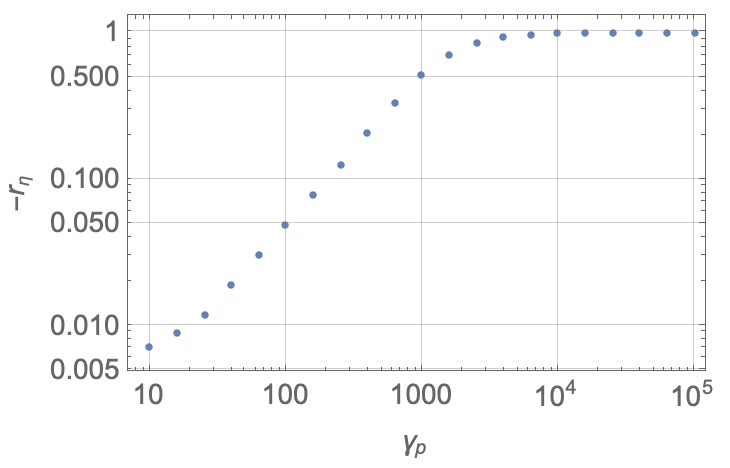}
  \caption{$r_\eta$ of the ${\rm TEM}_{11}$-mode as a function of the $\gamma_p$, where we assume the parameter of the DLW $a=0.5~[{\rm mm}]$, $\epsilon_r=3.75$, $r_{ab}=0.7613$.}
  \label{fig:mreta-gp}
\end{figure}

The factor
$(\gamma_p^2((1-\beta_p \beta_b) + (\beta_b - \beta_p)r_\eta^{(11)}))^{-1}$
in Eq.~\eqref{eq:zetamin}
comes from the missmatch between the
phase velocity of the deflecting mode $\beta_p$ and the velocity of the bunch $\beta_b$.
Fig.~\ref{fig:mreta-gp} shows the $r_\eta^{(11)}$ for the DLW with the vacuum-hole radius $a=0.5~[{\rm mm}]$ shown in Tab.~\ref{tab:dimension} for different $\gamma_p$.
We found $r_\eta^{(11)}$ is almost $-1$ in the region $\gamma_p>10^3$
and rapidly approaching to $0$ in the region $\gamma_p<10^3$,
then in the $v_p=c$ limit the impedance of the deflecting mode approaching to
the vacuum impedance with opposite sign,
and in the $v_p\ll c$ limit the deflecting field becomes TM-like mode, $|E_z|\gg Z_0 |H_z|$.

\begin{figure}[!htb]
  \centering
  \includegraphics*[height=5cm, bb= 0 0 353 244]{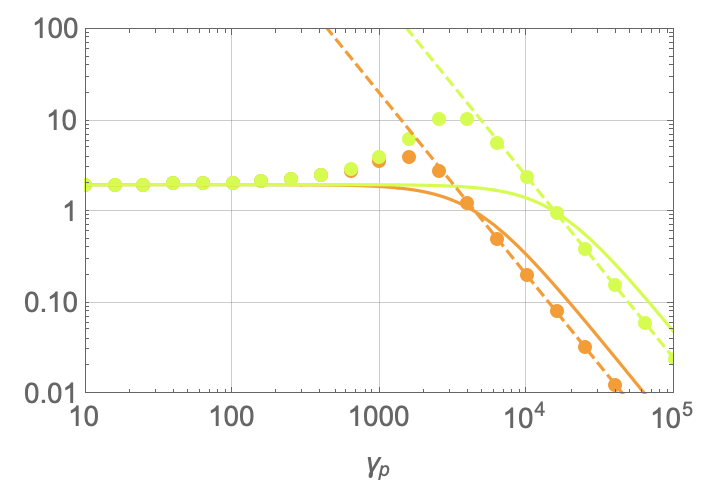}
  \caption{The $\gamma_p$ distribution of the fuction $(\gamma_p^2((1-\beta_p \beta_b) + (\beta_b - \beta_p)r_\eta^{(11)}))^{-1}$, the dots denote the
  result using the $\gamma_p$ distribution of the $r_\eta^{(11)}$ shown in Fig.~\ref{fig:mreta-gp}.
  The solid curves denote the limit of $r_\eta=0$, while the dashed curves
  denote the limit of $r_\eta=-1$. The orange and green color denote
  the beam energy $cp=2.3,\ 8~[{\rm GeV}]$, respectively.}
  \label{fig:fbeta-gp}
\end{figure}

The plot in Fig.~\ref{fig:fbeta-gp} shows the $\gamma_p$ dependence of
the factor $(\gamma_p^2((1-\beta_p \beta_b) + (\beta_b - \beta_p)r_\eta))^{-1}$
using $r_\eta^{(11)}(\gamma_p)$ shown in Fig.~\ref{fig:mreta-gp}.
When the phase velocity of the deflecting mode is approaching to the
bunch velocity, the factor can be up to $2,\ 10$ for the case of
the beam energy $cp=2.3,\ 8~[{\rm GeV}]$, respectively.
Then in the tuning of $\beta_p$ by the fine-tuning of
the frequency of the injected THz, it is better to set $\gamma_p<10^3$ for the optimum measurement.
As the peak of the factor is linearly proportional to the beam energy,
in the high energy limit this factor can limit the time resolution of the deflector.
Interestingly, once the condition $1>\beta_p>\beta_b$ satisfied, the factor becomes strong suppression and the time resolution is enhanced.

\subsection{Fraction of the wakefield effect to the deflecting mode}

In this section, we show the result of the calculation of Eq.~\eqref{eq:cpteG-sum}
using the parameter shown in Tab.~\ref{tab:dimension}.
To calculate the short-range wakefield, we will consider
up to $M$-th mode for ${\rm TEM}_{nm}$ satisfying $k_0^{(nM)}\sigma_z<1$,
which naturally introduces the effect of the bunch length in our model.
Tab.~\ref{tab:M} shows the number of modes we use to calculate the net contribution of the short-range wakefield to the Gaussian bunch.
In our analysis, we calculate the eigenfrequency of each configuration of DLW shown in Tab.~\ref{tab:dimension} up to $k_0<2.2\times 10^7[{\rm m^{-1}}]$, which
correspond to the bunch length of $\sigma_t=0.15~[{\rm fs}]$.

\begin{table}[htb]
  \begin{tabular}{ccccccc}\hline
      & \multicolumn{2}{c}{$a=0.5~[{\rm mm}]$ }& \multicolumn{2}{c}{$a=0.4~[{\rm mm}]$ }& \multicolumn{2}{c}{$a=0.3~[{\rm mm}]$ }\\ \hline
    $\sigma_t~[{\rm fs}]$ & $n=0$ & $n=1$ & $n=0$ & $n=1$& $n=0$ & $n=1$ \\ \hline
     $100$ & $6$ & $6$ & $3$ & $3$ & $2$ & $2$\\
     $10$   & $49$ & $55$ & $23$ & $23$ & $15$ & $15$\\
     $1$    & $422$ & $429$ & $198$ & $198$ & $131$ & $130$\\
     $0.15$   & $2433$ & $2509$ & $1088$ & $1146$ & $752$ & $751$\\ \hline
  \end{tabular}
  \caption{The maximum number of radial higher order mode $M$ depending on the bunch length $\sigma_t~[\rm fs]$ for each DLW shown in Tab.~\ref{tab:dimension}.}
  \label{tab:M}
\end{table}

In the calculation of the short-range wakefield, we need to take into account the factor $(1-v_g^{(nm)}/v_b)^{-1}$ for the higher-order mode
$m\leq M$.
For simplicity, we use the fitting result of the frequency distribution of the
group velocity as
$(1-v_g/c)^{-1}=1.72 + 7.16 \times 10^{-9}k_0^2$, where
we perform fitting in the region from $k_0=10^{3}~[{\rm m^{-1}}]$
to $k_0=10^{7}~[{\rm m^{-1}}]$.
Since for the mode of small wavelength $\lambda\ll a$ the DLW can be seen as
free space, the group velocity approaches to the speed of light.

\begin{figure}[!htb]
  \centering
  \includegraphics*[height=6cm, bb= 0 0 364 228]{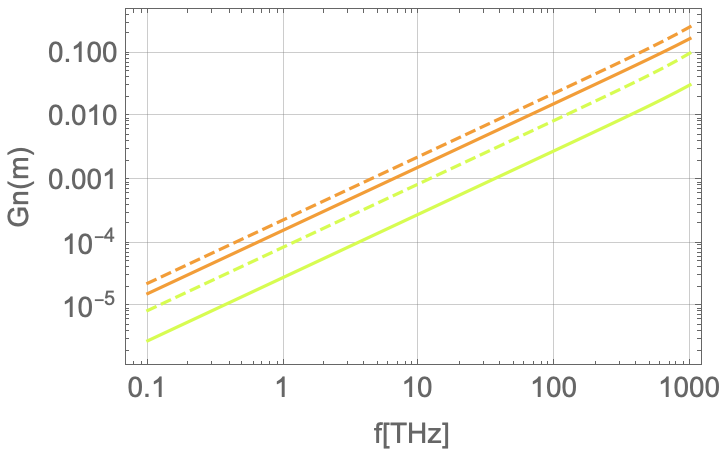}
  \caption{
  The frequency dependence of the $G_n$ for the DLW of $a=0.5$ shown in Tab.~\ref{tab:dimension}.
  The orange lines denote the result for $n=0$, the green lines denote
  the result for $n=1$.
  The solid line denote the result for $r_0=0.1~[{\rm mm}]$, the dashed line denote the
  result for $r_0=0.2~[{\rm mm}]$.
  We use the beam size of $\sigma_r=0.1~[{\rm mm}]$.
  }\label{fig:Gn-fthz}
\end{figure}

The form-factor $G_n(\alpha_1^{(nm)}\sigma_r,r_0/\sigma_r)$
takes into account the effect of the beam size and the particle loss
by the vacuum hole of the DLW.
Fig.~\ref{fig:Gn-fthz} shows the frequency dependence of the $G_n$
for the DLW with the vacuum hole radius $a=0.5~[{\rm mm}]$ shown in
Tab.~\ref{tab:dimension}.
Using the parameter fit, we obtain the scaling as $G_n\propto f^{x}$,
where $x=1.038,\ 1.058$ for $n=0$ and $r_0=0.1,\ 0.2~[{\rm mm}]$, and
$x=1.059,\ 1.079$ for $n=1$ and $r_0=0.1,\ 0.2~[{\rm mm}]$, respectively.

\begin{figure}[!htb]
  \centering
  \includegraphics*[height=6cm, bb= 0 0 364 237]{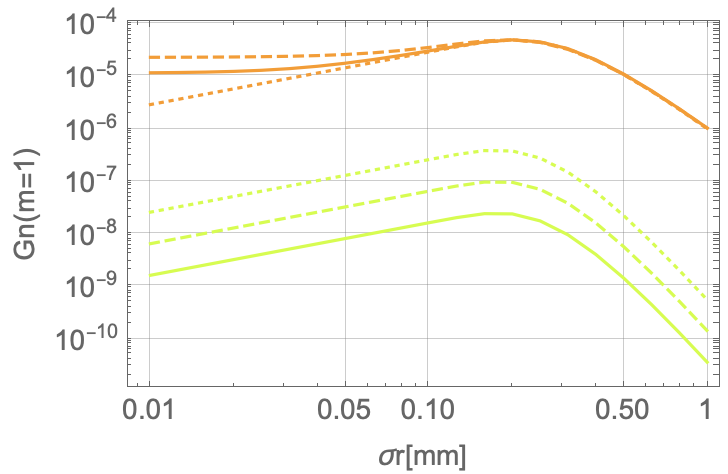}
  \caption{
  The beam size dependence of the $G_n$ for the mode of $(n,m)=(0,1),\ (1,1)$
  in the DLW of $a=0.5$ shown in Tab.~\ref{tab:dimension}.
  The orange lines denote the result for $n=0$, the green lines denote
  the result for $n=1$.
  The solid line denote the result for $r_0=0.05~[{\rm mm}]$, the dashed line denote the
  result for $r_0=0.1~[{\rm mm}]$, the dotted line denote the
  result for $r_0=0.2~[{\rm mm}]$.
  }
  \label{fig:Gnm1-sr}
\end{figure}

\begin{figure}[!htb]
  \centering
  \includegraphics*[height=6cm, bb= 0 0 367 233]{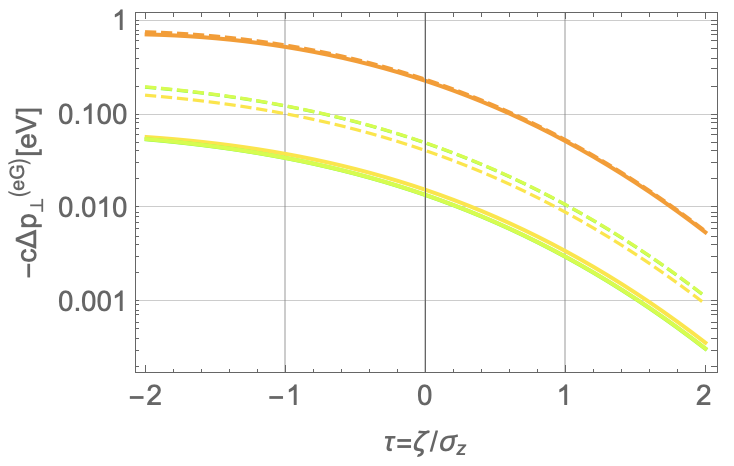}
  \caption{
  The momentum kick by the wakefield $\Delta p_\perp^{\rm (eG)}(\tau)$
  for the DLW of $a=0.5~[{\rm mm}]$.
  The orange lines denote the results for bunch length $\sigma_t=1~[{\rm fs}]$,
  the yellow lines denote ones for $\sigma_t=10~[{\rm fs}]$,
  and the green lines denote ones for $\sigma_t=100~[{\rm fs}]$.
  The solid lines denote the results for the offset $r_0=0.1~[{\rm mm}]$,
  and the dashed lines denote ones for the offset $r_0=0.2~[{\rm mm}]$.
  The horizontal axis denotes the longitudinal coorinate $\zeta$
  normalized by the bunch length $\sigma_z$.
  We assume the bunch with beam size
  $\sigma_r=a~[{\rm mm}]$, the bunch charge $Q=1~[{\rm pC}]$,
  and the length of DLW $L_d=10~[{\rm mm}]$.
  }
  \label{fig:pt-tau2}
\end{figure}

The transverse
momentum change from the RF and the wakefield depend on the relative position
in the bunch $\zeta=\tau \sigma_z$.
Using Eq.\eqref{eq:cprf} and
Eq.~\eqref{eq:cpteG-sum} we can calculate the
net transverse momentum change at the position $\tau$ as
\begin{eqnarray}
  c\Delta p_{\perp}(\tau)=
  c\Delta p_{\perp}^{\rm RF}(\tau)+c\Delta p_\perp^{({\rm eG})}(\tau).
\end{eqnarray}
Fig.~\ref{fig:pt-tau2} shows $-c\Delta p_{\perp}^{\rm (eG)}(\tau)/e$,
the deflecting momentum for the DLW of $a=0.5~[{\rm mm}]$
when the bunch charge is $Q=1~[{\rm pC}]$
for the bunch length $\sigma_t=1,\ 10,\ 100~[{\rm mm}]$
and the offset $r_0=0.1,\ 0.2~[{\rm mm}]$,
assuming the beam size is chosen as $\sigma_r=0.5~[{\rm mm}]$.

The effect of the wakefield becomes negligible when
the condition $\Delta p_{\perp}^{({\rm eG})}(\tau)\gg \Delta p_{\perp}^{\rm RF}(\tau)$ is satisfied.
Then we calculate the ratio of the two contributions as
\begin{eqnarray}
  r_{\rm w}(\tau) &=&
  \frac{\Delta p_{\perp}^{({\rm eG})}(\tau)}{\Delta p_{\perp}^{\rm RF}(\tau)},
  \nn\\
  &=&
  -\sum_{n=0}^N\sum_{m=1}^M
  \frac{Q}{2(1-v_g^{(nm)}/v_b)\tilde{U}^{(nm)}E_\perp^{\rm RF}}
  \frac{1}{
  \gamma_b k_z^{(11)} \sigma_z}
  G_n(\alpha_1^{(nm)}\sigma_r,r_0/\sigma_r)
  \frac{F_e(\tau,k_z^{(nm)} \sigma_z)}{\tau}.\label{eq:rw-tau}
\end{eqnarray}

Below we show the ratio $r_{\rm w}(\tau)$ at $\tau = -1$
which is around the bunch tail and has
larger wakefield effect than the head of the bunch.
We assume the bunch charge $Q=1~[{\rm pC}]$, and the dimension of the
DLW shown in Tab.~\ref{tab:dimension}, and beam energy $\gamma_b=10^4$.
We assume the peak gradient of the deflecting field
is calculated by the shunt impedance $Z_\perp$ shown in Tab.~\ref{tab:dimension}
considering
the case for the THz pulse with the energy of $10~[{\rm mJ}]$ and $10$ cycles.

\begin{figure}[!htb]
    \centering
    \includegraphics*[height=6cm, bb= 0 0 367 233]{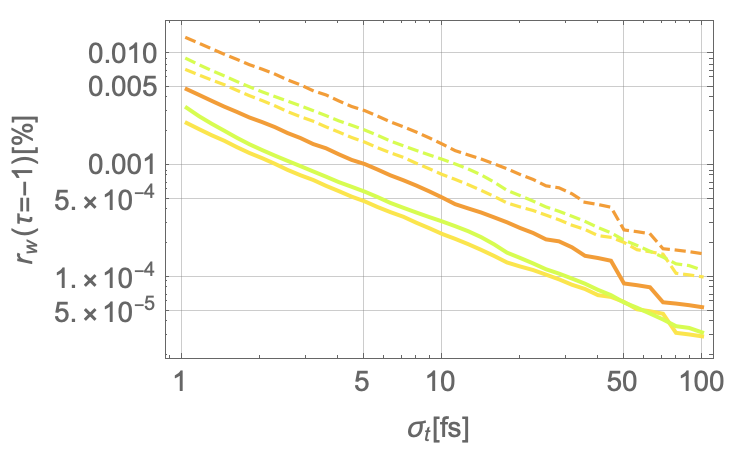}
    \caption{The fraction of the momentum kick from the wakefield to the one from the RF at the bunch tail $\tau=-1$ for $Q=1~[{\rm pC}]$ in terms of $\sigma_t~[{\rm fs}]$.
    The orange lines denote the result of $a=0.3~[{\rm mm}]$,
    and the yellow lines denote the result of $a=0.4~[{\rm mm}]$,
    and the green lines denote the result of $a=0.5~[{\rm mm}]$.
    The solid lines denote the result of $r_0=0.1~[{\rm mm}]$,
    and the dashed lines denote the result of $r_0=0.2~[{\rm mm}]$.
    }\label{fig:rw-tau2m1-stfs}
\end{figure}

Fig.~\ref{fig:rw-tau2m1-stfs} shows the ratio in terms of
$\sigma_t$ up to the bunch length of $\sigma_t=1~[{\rm fs}]$
for $a=0.3,\ 0.4,\ 0.5~[{\rm mm}]$, and $r_0=0.1,\ 0.2~[{\rm mm}]$.
Here the beam size is chosen as $\sigma_r=a$.
Considering the case of $M=1$ and $k_z^{(n1)}\ll1$,
the Eq.~\eqref{eq:rw-tau} does not depend on the bunch length $\sigma_t$.
Due to the increasing higher-order modes for the shorter bunch length,
the wakefield effect becomes larger than the shorter bunch length case.
In the plots shown in Fig.~\ref{fig:rw-tau2m1-stfs} we roughly obtain
$r_w\propto \sigma_t^{-1}$.

\begin{figure}[!htb]
  \centering
  \includegraphics*[height=6cm, bb= 0 0 367 233]{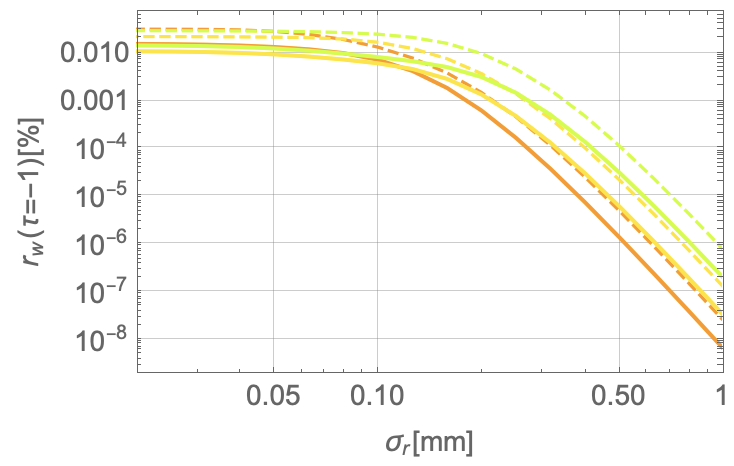}
  \caption{The fraction of the momentum kick from the wakefield to the one from the RF at the bunch tail $\tau=-1$ for $Q=1~[{\rm pC}]$ in terms of $\sigma_r~[{\rm mm}]$.
  The orange lines denote the result of $a=0.3~[{\rm mm}]$,
  and the yellow lines denote the result of $a=0.4~[{\rm mm}]$,
  and the green lines denote the result of $a=0.5~[{\rm mm}]$.
  The solid lines denote the result of $r_0=0.1~[{\rm mm}]$,
  and the dashed lines denote the result of $r_0=0.2~[{\rm mm}]$.
  }\label{fig:rw-tau2m1-srmm}
\end{figure}

Fig.~\ref{fig:rw-tau2m1-srmm} shows the ratio in terms of the beam size
$\sigma_r$ for $a=0.3,\ 0.4,\ 0.5~[{\rm mm}]$, and $r_0=0.1,\ 0.2~[{\rm mm}]$.
We assume the bunch length $\sigma_t=100~[{\rm fs}]$.
The reduction of the ratio for larger beam size $\sigma_r>a$ comes from
the beam loss by the beam hole.
For the smaller beam size compared with the beam hole $\sigma_r<a$,
the ratio is independent of the beam size.

\begin{figure}[!htb]
  \centering
  \includegraphics*[height=6cm, bb= 0 0 367 233]{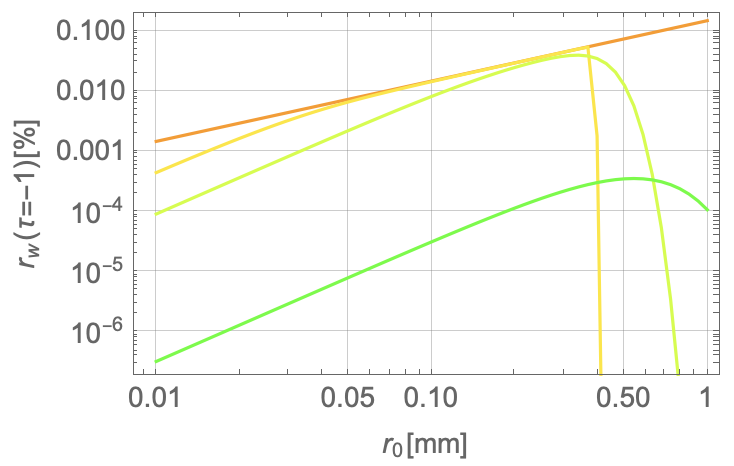}
  \caption{The fraction of the momentum kick from the wakefield to the one from the RF
  in terms of $r_0~[{\rm mm}]$ at the bunch tail $\tau=-1$ for $Q=1~[{\rm pC}]$ and the DLW with the beam hole radius $a=0.5~[{\rm mm}]$.
  The orange line denotes the result of $\sigma_r=0~[{\rm mm}]$,
  and the yellow line denotes the result of $\sigma_r=0.02~[{\rm mm}]$,
  and the light green line denotes the result of $\sigma_r=0.1~[{\rm mm}]$,
  and the green line denotes the result of $\sigma_r=0.5~[{\rm mm}]$.
  }\label{fig:rw-tau2m1-r0mm}
\end{figure}

Fig.~\ref{fig:rw-tau2m1-r0mm} shows the ratio in terms of the offset
$r_0$ for different beam sizes $\sigma_r=0,\ 0.02,\ 0.1,\ 0.5~[{\rm mm}]$
in the DLW with vacuum hole radius $a=0.5~[{\rm mm}]$.
We assume the bunch length $\sigma_t=100~[{\rm fs}]$.
In the limit line charge, $\sigma_r\to 0$, the ratio is
linearly proportional to the offset.
When the offset is larger than the beam size, $r_0>\sigma_r$,
the ratio proportional to the square of the offset.

\section{Conclusion}\label{sec:conclusion}

We investigate the RF deflector in the THz regime
to measure the ultrashort bunch with a length of ${\cal O}(1)~[{\rm fs}]$
and energy of GeV scale
used or to be used in the FEL facilities.
We analytically and numerically investigate the application of
the deflecting mode in
the dielectric-lined waveguide (DLW) structure as the deflector.
We found that for the sub-THz frequency
the DLW provides
fine time resolution of ${\cal O}(0.1)~[{\rm fs}]$
for the high energy beam with energy of several GeV,
keeping the wakefield effect less than $1~[\%]$
when $Q=1~[{\rm pC}]$.

We provide the formulae to design the DLW as a deflector
and to estimate the short-range wakefield to the Gaussian bunch
in a self-consistent manner.
By calculating the eigenfrequency we estimate the effect of the
fabrication error to the frequency fluctuation.
Considering the typical error of the dimension of the diameter of the DLW as
$\delta b=100~[{\rm \mu m}]$, we calculate the frequency fluctuation of the
deflecting mode in the fabrication.
The formulae for the shunt impedance and the group velocity are also given,
which provides the deflecting voltage
of the RF and the required pulse length of THz.

As an example, we investigate the time resolution and the
fraction of the momentum kick from the wakefield
in the typical parameter of the FEL facility, $pc=2.3,\ 8~[{\rm GeV}]$,
for the DLW supporting $0.2,\ 0.4,\ 0.6~[{\rm THz}]$.
Then we found the ideal time resolution can reach
order ${\cal O}(0.1)~[{\rm fs}]$
when the injected THz with the energy of $10~[{\rm mJ}]$ and 10 cycles per pulse.

Using our model which expresses the momentum kick from by
all higher modes generated by the particle in the bunch with
UV cutoff determined by the bunch length,
we compare the wakefield contribution at the tail of the bunch
to the RF contribution and numericaly obtain the following scaling laws.
The ratio of the momentum kick by the wakefield to the RF contribution is
linearly proportional to the bunch charge and the length of the deflector;
the ratio is proportional to the offset when the offset is larger than the
beam size, and to the square of the offset when the offset is smaller than
the beam size.
As long as the beam size is smaller than the vacuum hole radius of DLW
the ratio rarely depends on the beam size.
The ratio is inversely proportional to the bunch length.

\appendix

\section{The eigenmode analysis in the DLW}\label{sec:eigen}

The eigenmode analysis in the dielectric cylindrical waveguide has been investigated in the literature~\cite{cook2009generation} and the references therein.

We consider the metallic tube partially filled by the dielectric material with a permittivity of $\epsilon_r \epsilon_0$ which has a vacuum hole with permittivity of $\epsilon_0$.
The radius of the metallic tube is $b$ and one of the vacuum holes is $a$.
We will denote the vacuum region $(0\leq r\leq a)$ as subscript 1 and the dielectric region
$(a<r\leq b)$ as subscript 2, e.g. the radial wavenumber in the vacuum region, $k_{r1}$, and in the dielectric region, $k_{r2}$.

The electromagnetic field propagating inside the DLW with angular frequency $\omega$ and the propagating constant $k_z$ can be obtained from Helmholtz equation as eigenmodes with $n$ azimuthal waves and $m$ radial waves.
As for $n=0$ the $E_z$ and $H_z$ components are decoupled, then we can define the pure ${\rm TM}_{0m}$-mode for solutions with $H_z=0$ and the pure ${\rm TE}_{0m}$-mode for solutions with $E_z=0$.
On the other hand, as long as $k_z>0$ solutions with $n>0$ have non-zero $E_z$ and $H_z$ components, which are called ${\rm TEM}_{nm}$-mode.

Considering the case that the phase velocity $v_p$ is matching with the beam velocity $v_b<c$, the longitudinal wavenumber $k_z$ should be larger than the wavenumber in the free space $k_0$, namely $k_0 < k_z$.
Then the radial wavenumber in both region becomes
\begin{eqnarray}
  k_{r1}^2 &=& k_0^2 - k_z^2 \equiv - \alpha_1^2 <0,\\
  k_{r2}^2 &=& \epsilon_r k_0^2 - k_z^2 \equiv k_2^2>0.
\end{eqnarray}

The uniformity along the beam axis and Maxwell equation lead to the relation between the axial field components and others as follows:
\begin{eqnarray}
 {\vec E}_{ti} &=& \frac{i}{k_{ri}^2}\left[-k_z \vec\nabla_t E_{zi} + \frac{\w}{c} \hat z \times
\vec \nabla_t (Z_0 H_{zi})\right],\\
Z_0\vec{H}_{ti} &=& - \frac{i}{k_{ri}^2} \left[k_z \vec{\nabla}_t(Z_0 H_{zi})
+ \frac{\w}{c} \bm{\hat z}\times \vec \nabla_t E_{zi}\right],
\end{eqnarray}
where $Z_0\simeq377~[{\rm \Omega}]$ is the wave impedance in the vacuum.
Then we can reduce Eq.~\eqref{eq:helm-e} and Eq.~\eqref{eq:helm-h} into
\begin{eqnarray}
  (\nabla^2_T + k_{ri}^2 ) E_{zi}(r,\theta)  &=& 0,\\
  (\nabla^2_T + k_{ri}^2 ) H_{zi}(r,\theta)  &=& 0.
\end{eqnarray}

As $k_{r1}^2<0$ the radial dependence in the vacuum region should be
the linear combination of the modified Bessel functions of the first and second kinds, $I_n(\alpha_1 r)$ and $K_n(\alpha_1 r)$.
Due to the singularity at the origin, $\lim_{r\to \infty}K_n(\alpha_1 r)=\infty$,
only $I_n(\alpha_1 r)$ is enought.
As $k_{r1}^2>0$ in the dielectric region, the radial dependence is the linear combination of the Bessel functions of the first and second kinds, $J_n(k_2 r)$
and $Y_n(k_1 r)$.

\begin{eqnarray}
  E_{z1} &=& A_1 I_n(\alpha_1 r)\cos(n\phi)e^{i\psi(z,t)},\\
  H_{z1} &=& B_1 I_n(\alpha_1 r)\sin(n\phi)e^{i\psi(z,t)},\\
  E_{z2} &=& \left(A_2 J_n(k_2 r) + A_3 Y_{n}(k_2 r)\right)\cos(n\phi)e^{i\psi(z,t)},\\
  H_{z2} &=& \left(B_2 J_n(k_2 r) + B_3 Y_{n}(k_2 r)\right)\sin(n\phi)e^{i\psi(z,t)},
\end{eqnarray}
where $\psi(z,t) = \omega t - k_z z$.

We define the ratio of the amplitude of the axial electric field
and one of the axial magnetic field as
\begin{eqnarray}
  \eta=\frac{A_1}{B_1},\label{eq:eta}
\end{eqnarray}
which is
the impedance of the propagating mode in the DLW and calculated
by using Eq.~\eqref{eq:ephi} or Eq.~\eqref{eq:hphi}.

For latter simplicity of the arguments, we define $u=\alpha_1 a$, $v=k_2 a$
, $w=k_2 b$.
The boundary condition at the surface of the metal wall $E_{z2}|_{r=b}=0$ and $H_{r2}|_{r=b}=0$ give
\begin{eqnarray}
  A_3 &=& - \frac{J_n(w)}{Y_n(w)}A_2 \equiv A_{32}A_2\equiv A_{31}A_1,\\
  B_3 &=& - \frac{J_n'(w)}{Y_n'(w)}B_2 \equiv B_{32}B_2\equiv B_{31}B_1,
\end{eqnarray}
respectively.
We also define the ratio $A_{2}=A_{21}A_1$, $B_2=B_{21}B_1$.

The electromagnetic field in the vacuum region with azimuthal index $n$ is given as follows:
\begin{eqnarray}
  E_{z1}^n &=& E_0 I_n(\lambda_1) \cos{(n\phi)}\cos{\psi(z,t)},\\
  E_{r1}^n &=& -\gamma_p E_0
  R_n\left(\lambda_1,1,\beta_p r_\eta\right)\cos{(n\phi)}\sin{\psi(z,t)},\\
  E_{\phi1}^n &=& \beta_p \gamma_p E_0
  R_n\left(\lambda_1, r_\eta,1/\beta_p\right)\sin{(n\phi)}\sin{\psi(z,t)},\\
  H_{z1}^n &=& \frac{E_0}{Z_0} r_\eta I_n(\lambda_1) \sin{(n\phi)}\cos{\psi(z,t)},\\
  H_{r1}^n &=& - \gamma_p \frac{E_0}{Z_0}
  R_n\left(\lambda_1,r_\eta,\beta_p \right)\sin{(n\phi)}\sin{\psi(z,t)},\\
  H_{\phi1}^n &=& - \beta_p \gamma_p \frac{E_0}{Z_0}
  R_n\left(\lambda_1,1 ,r_\eta / \beta_p \right)\cos{(n\phi)}\sin{\psi(z,t)},
\end{eqnarray}
where $\lambda_1=\alpha_1 r$ and $r_\eta=Z_0/\eta$.
We use the relation $k_z/\alpha_1 = 1/\sqrt{1-\beta_p^2}\equiv \gamma_p$.
$Z_0$ is the impedance in free space, which relates to the permittivity $\epsilon_0$, the permiability $\mu_0$ and the velocity of light $c$ in free space as $\epsilon_0=1/cZ_0$
and $\mu_0=Z_0/c$.
For simplicity, we define the functions for radial dependence and longitudinal and time dependence as
\begin{eqnarray}
  R_n(\lambda,A,B) &=& A I_n'(\lambda) + B \frac{n I_n(\lambda)}{\lambda},\label{eq:rn}\\
  &\simeq&\begin{cases}
    \frac{A}{2}\lambda, & (n=0)\\
    \frac{A+B}{2}. & (n=1)
  \end{cases}
\end{eqnarray}
where we expand $R_n(r,A,B)$ around the beam axis $\lambda_1=\alpha_1 r(\ll 1)$.

The field in the region 2 ($a<r<b$) can be obtained as
\begin{eqnarray}
  E_{z2}^{n} &=& E_0 C_n(\lambda_2)\cos(n\phi) \cos{\psi},\\
  E_{r2}^n &=&  \frac{1}{\sqrt{\epsilon_r \beta_p^2-1}}
  E_0 R_{2n}(\lambda_2,1,\beta_p r_\eta)
  \cos(n\phi)\sin{\psi},\\
  E_{\phi2}^n &=& - \frac{1}{\sqrt{\epsilon_r \beta_p^2-1}}
  E_0 \bar{R}_{2n}(\lambda_2,\beta_p r_\eta,1)
  \sin(n\phi)\sin{\psi},\\
  H_{z2}^n &=& \frac{E_0}{Z_0} r_\eta D_n(\lambda_2) \sin(n\phi) \cos{\psi},\\
  H_{r2}^n &=& \frac{1}{\sqrt{\epsilon_r \beta_p^2-1}} \frac{E_0}{Z_0}
   \bar{R}_{2n}(\lambda_2,r_\eta,\epsilon_r \beta_p) \sin(n\phi) \sin{\psi},\\
  H_{\phi 2}^n &=& \frac{1}{\sqrt{\epsilon_r \beta_p^2-1}}\frac{E_0}{Z_0}
  R_{2n}(\lambda_2,\epsilon_r \beta_p,r_\eta) \cos(n\phi) \sin{\psi},
\end{eqnarray}
where $C_n(\lambda)=A_{21}J_n(\lambda)+A_{31}Y_n(\lambda)$,
$D_n(\lambda)=B_{21}J_n(\lambda)+B_{31}Y_n(\lambda)$,
$R_{2n}(\lambda,A,B)=AC_n'(\lambda)+B(nD_n(\lambda)/\lambda)$ and
$\bar{R}_{2n}(\lambda,A,B)=AD_n'(\lambda)+B(nC_n(\lambda)/\lambda)$.

As for the continuous condition at the boundary of the dielectric material and the vacuum, the tangential component of the electric field $\vec{E}_i$ and the magnetic field $\vec{H}_i$ should be same at the boundary.
For the longitudinal components,
$E_{z1}|_{r=a}=E_{z2}|_{r=a}$ and $H_{z1}|_{r=a}=H_{z2}|_{r=a}$ give
\begin{eqnarray}
  A_2 &=& \frac{I_n(u)Y_n(w)}{J_n(v)Y_n(w)-J_n(w)Y_n(v)}A_1 \equiv A_{21}A_1,\\
  B_2 &=& \frac{I_n(u)Y'_n(w)}{J_n(v)Y_n'(w)-J_n'(w)Y_n(v)}B_1 \equiv B_{21}B_1.
\end{eqnarray}
For the azimuthal components,
$E_{\phi1}|_{r=a}=E_{\phi2}|_{r=a}$
and $H_{\phi1}|_{r=a}=H_{\phi2}|_{r=a}$
give
\begin{eqnarray}
  \frac{B_1}{A_1} &=&
  -\frac{n k_z}{\omega \mu_0}
  \left[
  \frac{I_n(u)}{u^2} + \frac{1}{v^2}
  \left(A_{21} J_n(v)+A_{31}Y_n(v)\right)
  \right]\nn\\
  &&\left[
  \frac{I'_n(u)}{u} + \frac{1}{v}
  \left(B_{21} J_n'(v)+B_{31}Y_n'(v)\right)
  \right]^{-1}\label{eq:ephi}\\
  \frac{B_1}{A_1} &=&
  \frac{\omega \epsilon_0}{n k_z}
  \left[
  \frac{I_n(u)}{u^2} + \frac{1}{v^2}
  \left(B_{21}J_n(v) + B_{31} Y_n(v)\right)
  \right]^{-1}\nn\\
  &&\left[
  \frac{I_n'(u)}{u} + \frac{\epsilon_r}{v}
  \left(A_{21}J_n'(v) + A_{31}Y_n'(v)\right)
  \right].\label{eq:hphi}
\end{eqnarray}
Reducing \eqref{eq:ephi} and \eqref{eq:hphi}, we obtain the
dispersion relation of the eigenmodes in the DLW as,
\begin{eqnarray}
  &&f_{\rm disp}(k_0,k_z,a,r_{ab},\epsilon_r)\nn\\
  &=&
  \frac{k_0^2}{n^2 k_z^2}
  \left[
  \frac{I_n(u)}{u^2} + \frac{1}{v^2}
  \left(B_{21}J_n(v) + B_{31} Y_n(v)\right)
  \right]^{-1}\nn\\
  &&\left[
  \frac{I_n'(u)}{u} + \frac{\epsilon_r}{v}
  \left(A_{21}J_n'(v) + A_{31}Y_n'(v)\right)
  \right]\nn\\
  &&-
  \left[
  \frac{I_n(u)}{u^2} + \frac{1}{v^2}
  \left(A_{21} J_n(v)+A_{31}Y_n(v)\right)
  \right]\nn\\
  &&\left[
  \frac{I'_n(u)}{u} + \frac{1}{v}
  \left(B_{21} J_n'(v)+B_{31}Y_n'(v)\right)
  \right]^{-1},\label{eq:fdisp}
\end{eqnarray}
where we define the wavenumber in free space $k_0=\omega/c$ and
dimensionless parameter $r_{ab}=a/b$ to express the radius of the metal wall
as $b=a/r_{ab}$.
As for ${\rm TM}_{nm}$-mode, we obtain the resonant condition of the $m$-th solution
of $f_{\rm disp}(k_0,k_z,a,r_{ab},\epsilon_r)=0$, which constraints
the the frequency $f=ck_0/2\pi$ and the propagation constant $k_z=k_0/\sqrt{1-1/\gamma^2}$, the vacuum hole radius $a$,
the ratio of the radius of vacuum and dielectrics $r_{ab}=a/b$
, and relative permittivity $\epsilon_r$.

The energy per unit length $U=(1/2)\int dS (\epsilon |\vec{E}|^2 + \mu |\vec{H}|^2)$
and the power flux $P_z=\int dS (E_r H_\phi^* - E_\phi H_r^*)$ can be obtained as follows:
\begin{eqnarray}
  U &=& \frac{E_0^2}{2cZ_0}\frac{1}{\alpha_1^2}\Lambda_{u1}^n + \frac{\epsilon_r E_0^2}{2cZ_0}\frac{1}{k_2^2}\Lambda_{u2}^n,\nn\\
  &=& \frac{E_0^2}{2cZ_0}\frac{\gamma_p^2 \beta_p^2}{k_0^2}\Lambda_{u1}^n + \frac{E_0^2}{2cZ_0}\frac{\beta_p^2}{(\epsilon_r\beta_p^2 - 1)k_0^2}\Lambda_{u2}^n,\label{eq:uem}\\
  P_z &=& \frac{E_0^2}{Z_0}\frac{k_z^2}{\alpha_1^4}\Lambda_{p1}^n
  + \frac{E_0^2}{Z_0}\frac{k_z^2}{k_2^4}\Lambda_{p2}^n,\nn\\
  &=&\frac{E_0^2}{Z_0}\frac{\gamma_p^4\beta_p^2}{k_0^2}\Lambda_{p1}^n
  + \frac{E_0^2}{Z_0}\frac{\beta_p^2}{(\epsilon_r \beta_p^2 - 1)^2 k_0^2}\Lambda_{p2}^n\label{eq:pz}
\end{eqnarray}
where we define the dimensionless integral as
\begin{eqnarray}
\begin{split}
  \Lambda_{u1}^n &=
  \int_0^{2\pi}d\phi\int_0^u d\lambda_1 \lambda_1 \Big[
  (c_n^2 + r_\eta^2 s_n^2)I_n^2(\lambda_1)\nn\\
  &+ \gamma_p^2
  \Big(c_n^2(R^2_{n}(\lambda_1,1,\beta_p r_\eta) + R^2_{n}(\lambda_1,\beta_p,r_\eta))\nn\\
&+ s_n^2 (R^2_{n}(\lambda_1,\beta_p r_\eta,1) + R^2_{n}(\lambda_1,r_\eta,\beta_p))
\Big)
\Big],\nn\\
  \Lambda_{u2}^n &=
  \int_0^{2\pi}d\phi\int_v^w d\lambda_2 \lambda_2 \Big[
  (\epsilon_r c_n^2 C_n^2(\lambda_2)+r_\eta^2 s_n^2 D_n^2(\lambda_2) )\nn\\
  &+ \frac{1}{\epsilon_r\beta_p^2-1}
  \Big(
  c_n^2(\epsilon_r R^2_{2n}(\lambda_2,1,\beta_p r_\eta) + R^2_{2n}(\lambda_2,\epsilon_r \beta_p,r_\eta))\nn\\
  &+ s_n^2( \epsilon_r \bar{R}^2_{2n}(\lambda_2,\beta_p r_\eta,1) +
  \bar{R}^2_{2n}(\lambda_2,r_\eta,\epsilon_r \beta_p))
  \Big)
    \Big],\nn\\
    \Lambda_{p1}^n &=
    \int_0^{2\pi}d\phi\int_0^u d\lambda_1 \lambda_1 \Big[
    c_n^2 R_n(\lambda_1,1,\beta_p r_\eta)R_n(\lambda_1,\beta_p,r_\eta)\nn\\
    &+ s_n^2 R_n(\lambda_1,\beta_p r_\eta,1)R_n(\lambda_1,r_\eta,\beta_p)
    \Big],\nn\\
    \Lambda_{p2}^n &=
    \int_0^{2\pi}d\phi\int_0^u d\lambda_2 \lambda_2 \Big[
    c_n^2 R_{2n}(\lambda_2,1,\beta_p r_\eta)R_{2n}(\lambda_2,\epsilon_r\beta_p,r_\eta)\nn\\
    &+ s_n^2 \bar{R}_{2n}(\lambda_2,\beta_p r_\eta,1)\bar{R}_{2n}(\lambda_2,r_\eta,\epsilon_r\beta_p)
      \Big].\nn
  \end{split}
\end{eqnarray}
Here, we use $s_n=\sin(n\phi)$ and $c_n=\cos(n\phi)$.

\section{Time resolution of the RF deflector}\label{sec:resolution}

Following the discussion in the literature~\cite{malyutin2014time} and the references therein,
we briefly describe the estimate of the time resolution of the RF deflector.

Considering the entrance of the deflector is at $s_0$, the exit of it is at $s_0'$, and the screen is at $s$ with distance $L\equiv s-s_0\gg L_d \equiv s_0' - s_0$,
the particle of electric charge e at displacement $\zeta$ from bunch center change its horizontal position at the screen by
\begin{eqnarray}
  x(L,\zeta) = \frac{\Delta p_{\perp}}{p} L,
\end{eqnarray}
where $\Delta p_{\perp}$ is the net transverse momentum change while the bunch passes the deflector and $p$ is the magnitude of the momentum of the particle.

Considering the phase velocity of the propagating deflecting mode is $\beta_pc$ and
the velocity of the bunch is $\beta_b c<c$,
the particle in the bunch at $\zeta$ rides the THz phase $\psi(z,t)|_{t=(z-\zeta)/v_b} = -k_b \zeta + \delta k z$,
where $k_b=k_0/\beta_b$ and $\delta k=(\beta_b^{-1}-\beta_p^{-1})k_0$.

Then the particle experience the momentum kick from the deflecting mode,
\begin{eqnarray}
  c\Delta p_{\perp}^{\rm RF} &=& e E_{\perp} \gamma_p^2
  \left[
  (1-\beta_p \beta_b) + (\beta_b - \beta_p)r_\eta
  \right]
  \int_0^{L_d}dz (-\sin\psi(z,t))|_{t=(z-\zeta)/v_b},\nn\\
  &=& e E_{\perp} \gamma_p^2
  \left[
  (1-\beta_p \beta_b) + (\beta_b - \beta_p)r_\eta
  \right]\frac{-L_d}{\delta k}
   \left(
  \cos(k_b \zeta) - \cos(k_b\zeta - \delta k L_d)
  \right),\nn\\
  &\sim& e V_{\perp}
  \gamma_p^2
  \left[
  (1-\beta_p \beta_b) + (\beta_b - \beta_p)r_\eta
  \right]
  \left(
  k_b \zeta - \frac{1}{2}\delta k L_d
  \right),\label{eq:cptRFB2}
\end{eqnarray}
where $V_{\perp}=E_{\perp}^{\rm RF}L_d$ is the maximum deflecting voltage of the deflector.
Considering the synchronizing particle with $\beta_b = \beta_p$, the
momentum kick becomes
$c\Delta p_{\perp}^{\rm RF} =eV_\perp k_b \zeta$.

Assuming the RF contribution dominates the transverse momentum,
the slice at $\zeta$ in the bunch is projected on
the horizontal position at the screen,
\begin{eqnarray}
  x(L,\zeta) \sim \frac{eV_\perp}{pc}
  \gamma_p^2
  \left[
  (1-\beta_p \beta_b) + (\beta_b - \beta_p)r_\eta
  \right]
  \left(k_b \zeta - \frac{1}{2}\delta k L_d\right)L.
\end{eqnarray}

Considering the two slice at $\zeta,\ \zeta+\Delta \zeta$ in the bunch,
the condition to resolve them on the screen is
\begin{eqnarray}
  x(L,\zeta+\Delta \zeta) - x(L,\zeta) > x_{\rm rms}(s),
\end{eqnarray}
where the right-hand side denotes the bunch size on the screen without
RF.
Then we obtain
\begin{eqnarray}
  L\frac{1}{p}
    \frac{d \Delta p_\perp(\zeta)}{d\zeta}
  \Delta \zeta > x_{\rm rms}(s).\label{eq:resolution-condition}
\end{eqnarray}
Therefore we can obtain the minimum value of the $\Delta\zeta$ as
\begin{eqnarray}
  \Delta\zeta &>&  \frac{cp}{e V_\perp k_b\sin(\Delta \psi)}
  \frac{1}{\gamma_p^2((1-\beta_p \beta_b) + (\beta_b - \beta_p)r_\eta)}
  \frac{\epsilon_x}{x_{\rm rms}(s_0)\sin(\Delta \psi_x)}
  ,\label{eq:resolution}
\end{eqnarray}
where we use the relation
$L=\sqrt{\beta_x(s)\beta_x(s_0)}\sin(\Delta\psi_x)$
and
$x_{\rm rms}^2(s)=\beta_x(s) \epsilon_x$.

\section*{Acknowledgements}
We would like to acknowledge Prof.~Hiroyasu Ego for useful discussion.

\bibliography{ref}
\bibliographystyle{apsrev4-2}
\end{document}